**The Mass of Cosmic Rays of Ultra-High Energy**

A A Watson

School of Physics and Astronomy

University of Leeds, Leeds, LS2 9JT, UK

a.a.watson@leeds.ac.uk

**Abstract:** A review of several analyses is presented that forces the conclusion that the mass composition of the highest-energy cosmic rays is not proton-dominated. This deduction, combined with the use of a modern hadronic interaction model, should lead to a re-evaluation of the energy spectrum reported by the Telescope Collaboration that may well bring that measurement, and the corresponding one from the Pierre Auger Observatory, into better agreement.

1. **Introduction:** Knowledge of the mass composition and flux of ultra-high energy cosmic rays underpins efforts to understand the origin of these particles. Currently, the energy spectra reported by the Auger Collaboration and by the Telescope Array Collaboration are in strong disagreement above 10 EeV [1], even when fluxes from the same declination band are compared. As the latter group make assumptions about the primary mass in the derivation of their spectrum, it is important to address the question of what mass to adopt, both for energy estimation and for evaluation of the detection aperture near the threshold for detection where the collecting area varies rapidly with energy and is mass dependent [2]. Further, the Telescope Array Collaboration continues to use the QGSJet-II series of models of hadronic interactions, first introduced twenty years ago. These models have been superseded by others rooted in LHC data.

The idea that the highest-energy cosmic rays are dominantly protons has influenced thinking about their origin for 70 years. In 1955, the distinguished cosmic-ray and particle physicist, G Cocconi [3] wrote:

*"We remain with the dilemma: protons versus heavy nuclei. A clear-cut decision cannot be reached yet. I believe that up to the highest energies the protons are the most abundant in the primary cosmic rays. …………………… However, I must confess that a leak proof test of the protonic nature of the primaries at the highest energies does not exist. This is a very important problem. Experimentally it is quite a difficult problem."*

The final sentence has proved to be a major understatement.

Particles above 10 EeV were first observed by the MIT group [4]. The conclusion of Linsley and Scarsi about the primary mass [5], based on observations with the 3.3 $m^2$ muon detector of the Volcano Ranch array, was that the muon distribution above ~0.1 EeV implied that either protons or nuclei of the Fe group dominate, with arguments based on inclined showers appearing to favour protons [6]. The Volcano Ranch results on mass related to the range 0.1 EeV to 1 EeV and their detector was rather small. Muon detectors are expensive, and other methods have been developed to address the mass question at the highest energies.

That the proton idea became so embedded in the thinking of the cosmic-ray community is rather alarming. There has been an almost Orwellian-like groupthink, probably stimulated by the first direct measurements of the depth of shower maxima from the Fly's Eye experiment [7]. The conclusion of proton dominance made there, dependent as it was on the choice of hadronic model used to predict the evolution of shower maximum with energy, appeared to give weight to the proton paradigm and received more attention than counter-indications that took account of methods of particle acceleration.

The idea that the highest-energy particles might be of extragalactic origin, first discussed by Khristiansen and Kulikov [8], led Gerasimova and Rozenthal [9] to point out that nuclei with energies of $10^{16}$ eV/nucleon would be subject to photodisintegration in intergalactic space, and, with the photon fields assumed, to the claim that most heavy nuclei would disappear. However, Ginzburg and Syrovatskii [10] argued that the intergalactic photon density was a least an order of magnitude lower and concluded that a large fraction of heavy nuclei would survive, as is now commonly held to be the case. After the theory of diffusive shock acceleration became established, Hillas [11] pointed out that the highest energy that would be attained was proportional to Z, the atomic number of the nucleus of interest, rigidity being the key parameter.

Mass estimates that rely on the extrapolation of accelerator data have systematic uncertainties as, of course, do those based on astrophysical approaches. In both cases, the magnitude of the uncertainty is difficult to evaluate.

## 2. Inferences about Primary Mass from Measurements of the Depth of Shower Maximum

**2(a) Introduction:** The use of muon detectors to infer the primary mass has largely been superseded by determining the depth at which the energy loss of air showers reaches maximum, $X_{max}$, and comparing the evolution of this depth with energy, the Elongation Rate, with predictions based on models of hadronic interactions and on different choices of primary mass.

The Elongation Rate, a term introduced by Linsley [12], can be estimated robustly using analytical methods. Linsley also developed the Elongation Theorem [12], arguing that the limit to the Elongation Rate for a single species is $2.3(1\text{-B})X_o$ per decade, where $X_o$ is the radiation length in air and B the logarithmic derivative of the pion multiplicity with energy. With B= 0.2 [13] and $X_o = 37.7$ g cm$^{-2}$, this limit is 69 g cm$^{-2}$ per decade. The difference in $X_{max}$ between different nuclear masses, $A_1$ and $A_2$, is $(1\text{- B})X_o(\ln A_2 - \ln A_1)$, giving a proton-iron separation of ~120 g cm$^{-2}$ [14]. The upper limit to the Elongation Rate is ~87 g cm$^{-2}$ per decade, corresponding to a photon primary (or to a proton primary in which the leading particle from the first collision is a neutral pion).

From the Heitler-Matthews simplified treatment of shower development, the corresponding number derived for the Elongation Rate is 58 g cm$^{-2}$ per decade [15].

Predictions of the Elongation Rates from simulations made using different hadronic interaction models give values for proton and iron nuclei of ~54 and ~64 g cm$^{-2}$ per decade respectively (e.g. [16]).

Predictions of $X_{max}$ made using models of hadronic interactions have changed markedly over the years. In [7], the QCD Pomeron model was adopted, predicting $X_{max}$ for protons at 10 EeV as 780 g cm$^{-2}$. Corresponding numbers from the more recent EPOS-LHC and Sibyll 2.3c models are 805 and 810 g cm$^{-2}$ respectively [17] while for the EPOS-LHC-R model, recently developed by Pierog [18], the depth of maximum is predicted to be 820 g cm$^{-2}$. The detailed estimates of the distribution of different masses at a given energy, such as have been made by the Auger Collaboration [16, 17], depend on the choice of hadronic models at centre-of-mass energies up to ~30 greater than achieved at the LHC and so rely on extrapolations of parameters such as the cross-sections and inelasticities.

By contrast, any change in the mean mass with energy will be reflected in the Elongation Rate. It seems reasonable to conclude, both from simulations and from analytical methods, that the Elongation Rate for a pure composition is in the range (50 – 70) g cm$^{-2}$ per decade, with a separation of ~ (100 – 120) g cm$^{-2}$ between the depth of maxima of proton and iron nuclei at the same energy.

While estimates from simulations of the depth of shower maximum for proton primaries have changed by around 40 g cm$^{-2}$ over the last 30 years, the results from measurements have remained relatively stable. Indeed, the earliest estimate of $X_{max}$, derived from Chacaltaya data [19] using the constant intensity cut method [20], is (660 ± 40) g cm$^{-2}$ at ~0.1 EeV (using modern estimates of the primary energy), in good agreement with the direct measurements.

**2(b) Measurements of $X_{max}$ using fluorescence detectors:** Values of shower maxima measured using fluorescence detectors at the Auger Observatory [16] and at the Telescope Array [21] are shown in figure 1. Both datasets extend to about 40 EeV. The sensitivity of conclusions that might be drawn about primary mass to the choice of model is evident but a particular clear observational result, seen in the data from the Auger Observatory, is the break in the Elongation Rate at ~3 EeV (LH of figure 1). This break in the Elongation Rate provides strong evidence for a change to a heavier mean mass, unless the proton-air cross-section changes in an unexpected manner.

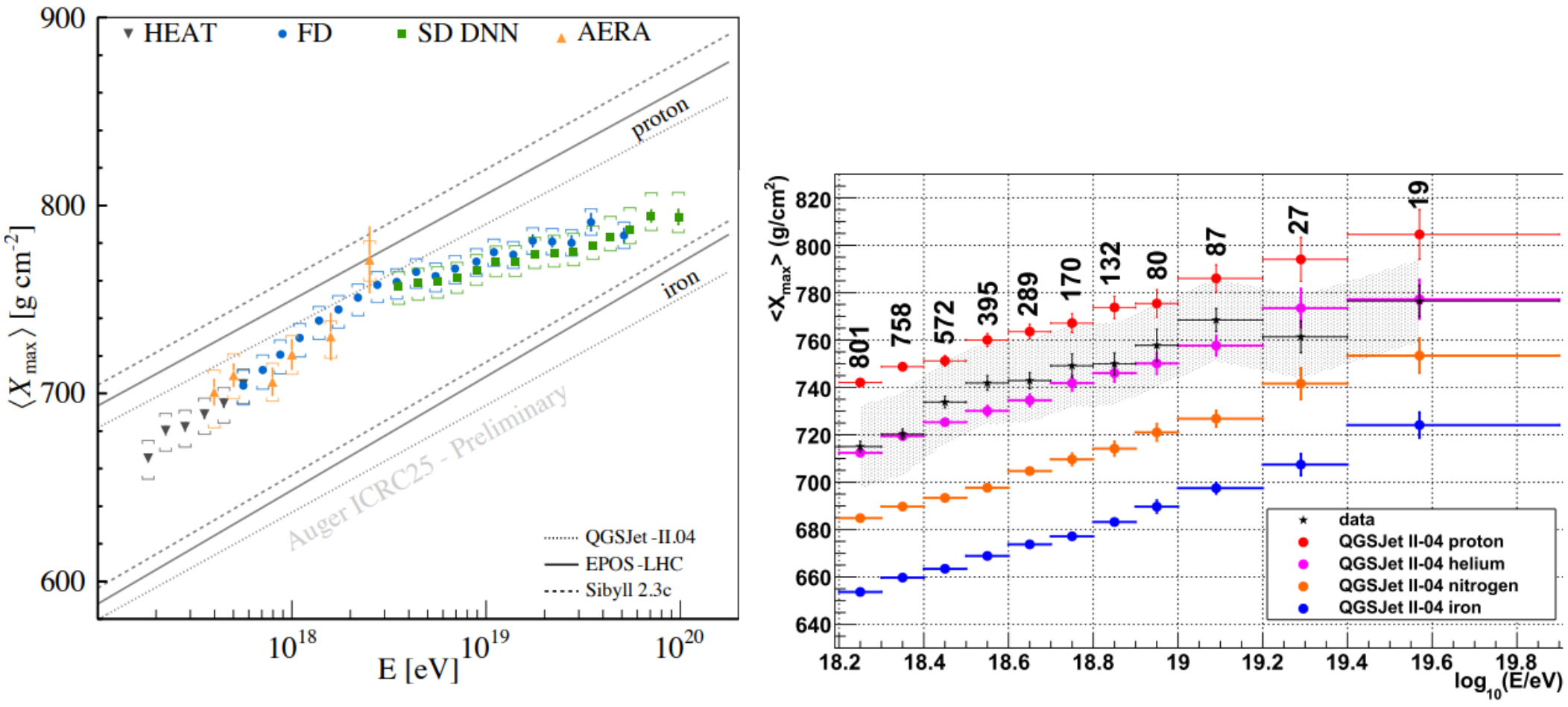

**Fig. 1.** LH plot: Measurements of $X_{max}$ by the Auger Collaboration [16] by a variety of methods: fluorescence measurements are HEAT and FD; RH plot: Measurements of $X_{max}$ by the Telescope Array Collaboration [21]. HEAT is the High-Elevation Auger Telescope while FD is the set of fluorescence detectors. Details of both instruments are given in [22].

Although not claimed by the Telescope Array Collaboration, there is also evidence in their data of a similar break (RH figure 1), as is shown in figure 2 [taken from 23] where the 11 data points from [21] have been re- plotted. There are 3330 events, of which 127 have energies above 10 EeV. The dashed line is a fit made assuming that the data can be described by a single line. That this fit is poor is evident by eye and is borne out by the reduced $\chi^2$ = 27.4/9 which corresponds to the probability of a single line fit being adequate as < $10^{-3}$.

Abbasi et al. [21], however, reported a reduced $\chi^2$ = 10.67/9 for their fit to a single line, a probability of an adequate fit of ~0.3. This value of $\chi^2$ can be reproduced if the widths of the energy bins are, unconventionally, included in the fitting procedure [23]. The measurements of $X_{max}$ reported from the Telescope Array are thus consistent with a break in the Elongation Rate and so with the mean mass increasing above 3 EeV.

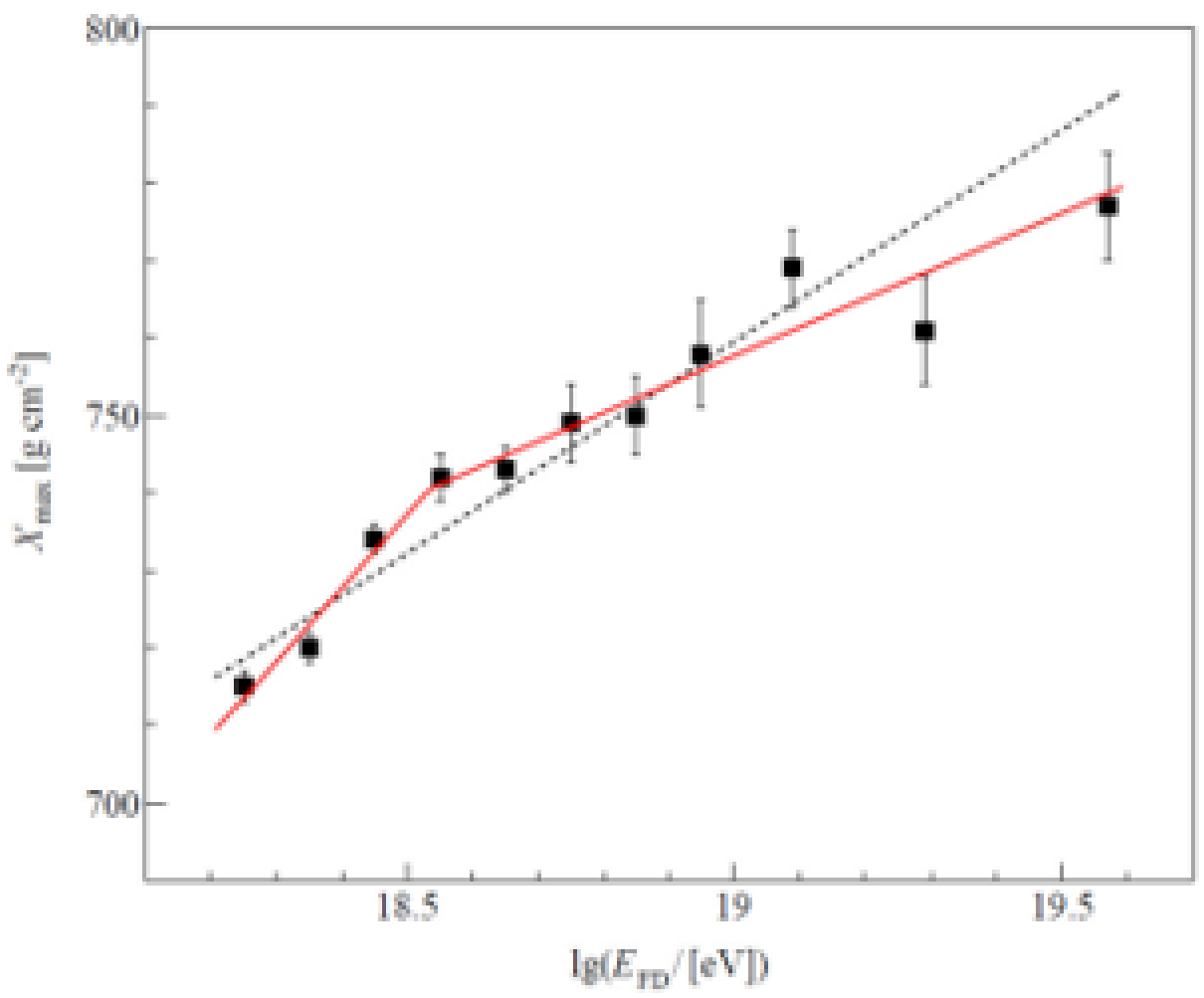

**Fig. 2.** $X_{max}$ vs Energy using data in [21, table 4]. The dashed line is a fit to the points with a single straight line (Elongation Rate = (54 ± 3) g cm$^{-2}$ per decade), while the full line is a fit with a break and with the best-fit straight lines found simultaneously (ER I = (95 ± 11) and ER II = (37 ± 6) g cm$^{-2}$ per decade). ERI and ERII are the Elongation Rates before and after the break energy.

The Telescope Array Collaboration have recently reported further measurements of $X_{max}$ made using the TALE and TAx4 fluorescence detectors [24, 25] which substantiate this conclusion. These new data are shown in figure 3, together with those of Abbasi et al. [21].

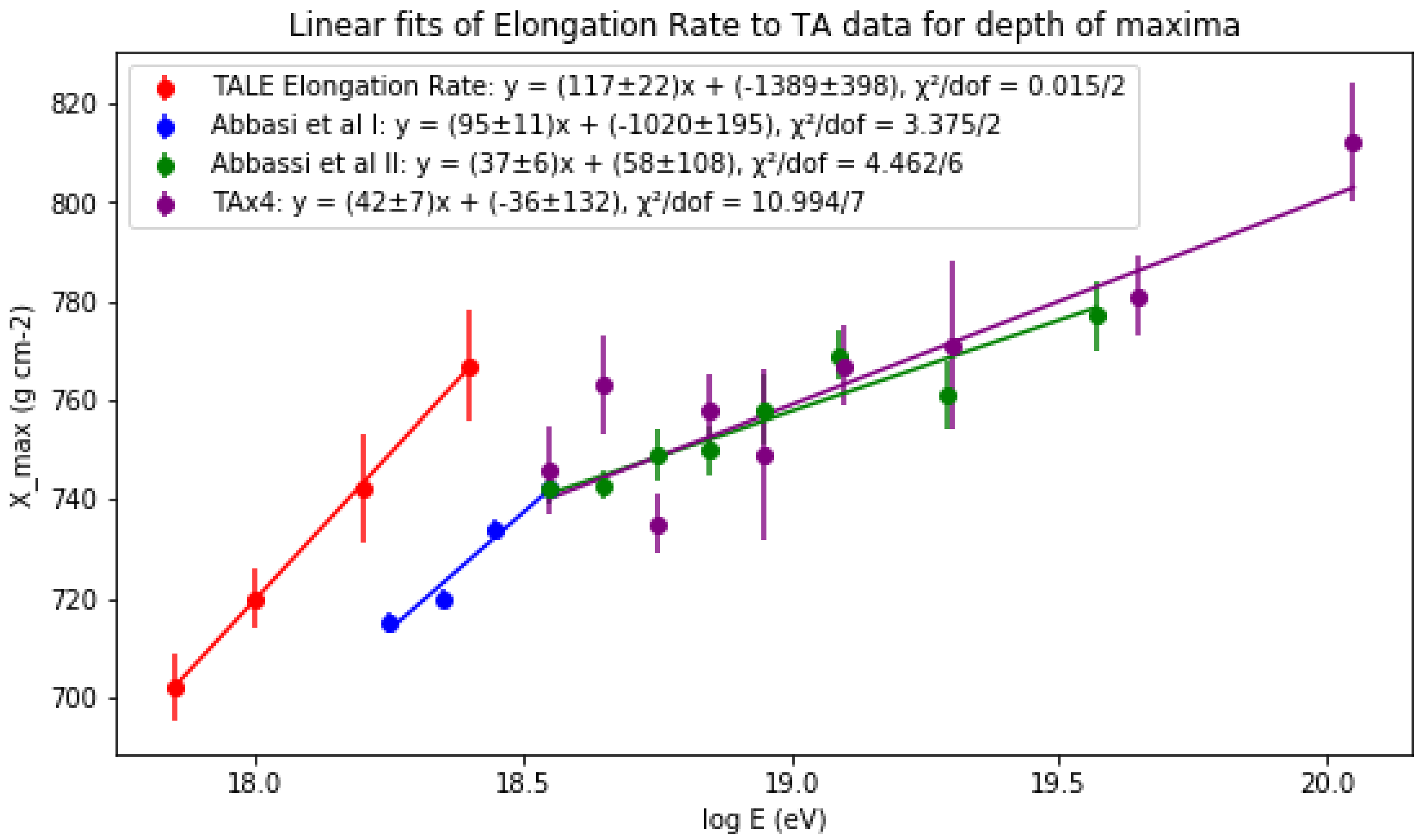

**Fig. 3.** Data on depth of shower maxima as reported by the Telescope Array [21, 24, 25]. ER is the Elongation rate; ERI and ERII are the values of the Elongation Rates from Abbasi et al. before and after ~3EeV (figure 2).

While there is evidence of a systematic difference between the TALE measurements [25] of $X_{max}$ and those reported in [21] in the overlap region below ~3 EeV, the two Elongation Rates are in reasonable agreement ((117 ± 22) and (95 ±11) g cm$^{-2}$ per decade respectively). Both sets of data are marked as 'preliminary' in [24, 25]: the uncertainties presently shown in the TALE plot [25], may be too large, as deduced from the lack of scatter of the four points around the straight-line fit.

The TAx4 data, based on 332 events of which 95 are above 10 EeV, are in good agreement with the results of Abbasi et al. [21]. The weighted average of the two Elongation Rates, above the break at ~3.5 EeV, is (39 ± 5) g cm$^{-2}$ per decade, well outwith the range (50 – 70) g cm$^{-2}$ per decade predicted for a pure species using either simulations or analytical methods, as discussed in section 2(a), and considerably smaller than the independent values of the Elongation Rate found by TALE [25] and Telescope Array [21], at lower energies, for which the weighted mean of the measurements is (99 ± 10) g cm$^{-2}$ per decade. The depth of shower maximum at 10 EeV from the sets of measurements [21, 24] is (760 ± 4) g cm$^{-2}$ which is out of line with the predictions for pure protons from EPOS-LHC of 805 g cm$^{-2}$, the smallest value derived from the more modern models (section 2(a)). Thus, both the Elongation Rate *and* the absolute value of $X_{max}$ provide evidence against proton-dominance at high energies.

The second moment of the $X_{max}$ distributions can also be compared and those from the two observatories are in good agreement. For both sets of measurements [17, 21], the sigmas of the $X_{max}$ distributions fall from ~60 g cm$^{-2}$ at 2 EeV to ~40 g cm$^{-2}$ at 10 EeV. Such changes are again indicative of the mean mass of the primary particles increasing with energy.

**2(c) Measurements of $X_{max}$ made using Water-Cherenkov Detectors:** Above ~40 EeV, the number of events for which $X_{max}$ can be measured directly is limited by the on-time of the fluorescence detectors that are only operable on clear moonless nights. By contrast, the on-time of the surface detectors of a shower array is close to 100%. Recently an approach has been developed to extract the depth of shower maximum from the time profile of the signals recorded at the 10 m$^2$ water-Cherenkov detectors of the Auger Observatory using deep-learning techniques [26], thus extending the energy range that can be explored with high statistics. These new measurements extend to 100 EeV and are shown in the left-hand plot in figure 1 (points marked SD DNN). The evolution of the mass towards a heavier composition is confirmed. This recent work has also exposed complexities in the evolution of the depth of shower maximum with energy which are not discussed here as they do not change the overall conclusion that the mean mass increases with energy.

**3**. **Inferences about primary mass from studies of arrival directions:** A different approach to gaining information about the primary mass at the highest energies has been developed by the Telescope Array Collaboration [27]. Their method is based on comparing the energy-dependent distribution of arrival directions with the expectations from a model for the origin of UHECR made with the reasonable assumption that the sources trace the large-scale structure of the Universe. In the case of negligible extra-galactic magnetic fields, they argue that the observations are consistent with a relatively heavy injected composition at ~10 EeV that becomes lighter up to ~100 EeV, while the composition at E >100 EeV is very heavy. The latter statement is claimed to be true even in the presence of the highest extra-galactic magnetic fields allowed experimentally, while the composition at lower energies can be light if a strong extra-galactic magnetic field is present. They conclude that protons are not dominant above 10 EeV.

**4. Inferences about primary mass from IceCube Observations:** It is well-known that the flux of the highest-energy neutrinos is lower if the primary beam is made up of nuclei heavier that protons [28, 29]. Recently, the IceCube Collaboration have reported a null result from their search for neutrinos with energies well-above 10 PeV [30]. This non-observation constrains the all-flavour neutrino flux at 1 EeV so that, if the cosmological evolution of the sources is comparable to or stronger than the star formation rate, the result disfavours the proton-only hypothesis. The IceCube Collaboration conclude that above 30 EeV, the proton fraction is constrained to be less than 70% at the 90% confidence level.

## 5. Inferences about primary mass from measurements of the energy spectrum:

A further argument favouring a heavy mass composition at the highest energies comes from an approach in which Neural Network techniques have been developed to improve analysis of data from the Telescope Array [31]. With this method, a mass-dependent bias in the energy estimates has been identified (figure 4). Using either the QGSJETII-04 or the SIBYLL 2.3d models of hadronic interactions, this bias is substantially reduced when iron nuclei rather than protons are selected as the primary particles.

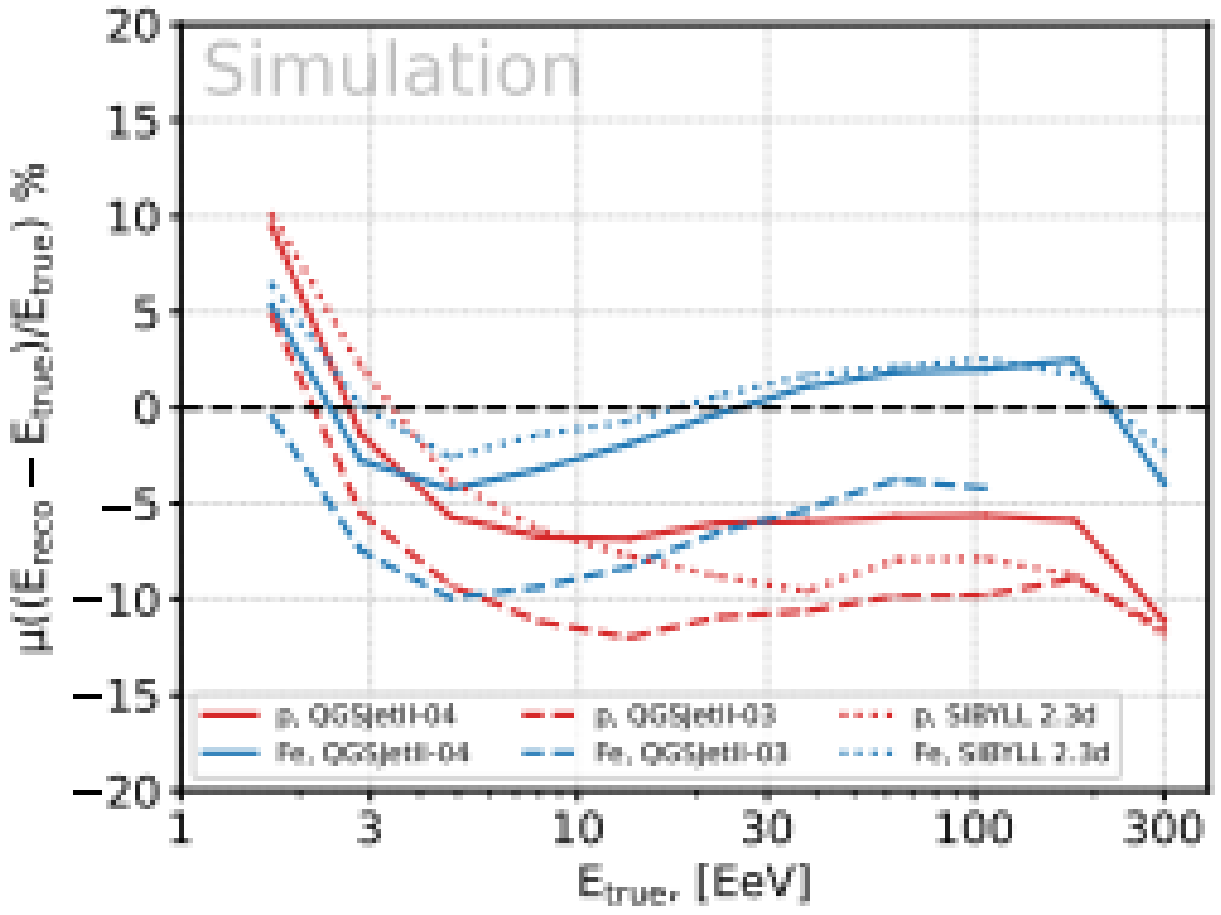

**Fig. 4.** Comparison of Deep Neural Network energy reconstruction biases across different hadronic interaction models [31]

**6. Conclusions:** Evidence from a variety of analyses has been presented that indicates that the mass of the highest energy cosmic rays is not proton-dominated. This conclusion should become the new 'groupthink'. While only by making assumptions about models of hadronic interactions can the contributions of each nuclear species be assessed, it seems clear that the conclusion of a dominantly protonic composition, as asserted most recently by the Telescope Collaboration [24], is weak. Recognition of this evidence, along with the adoption of a model tuned to LHC data, should prompt re-evaluation of the estimates of the flux of ultra-high energy cosmic rays reported by the Telescope Array. Such a re-evaluation is expected to bring those measurements into better accord with those from the Auger Collaboration. Certainly, this work should be carried out before attempting to explain the present differences in astrophysical terms. A review of the energy of the 244 EeV event reported by the Telescope Array Collaboration, which was assumed to have been initiated by a proton [32], should also be undertaken.

### Declaration of Competing Interests

The author declares that he has no known competing financial interests or personal relationships that could have appeared to influence the work reported in this paper.

### Acknowledgments

Antonio Bueno, Piera Ghia, Jim Matthews and Michael Unger are thanked for their comments on an early version of this paper. A referee is thanked for comments that have helped further improve it.